\documentclass[aps, twocolumn, showpacs, amsmath]{revtex4}
\usepackage{dcolumn}
\usepackage{bm}
\usepackage{graphicx}
\usepackage{color}
\usepackage{subfigure}
\usepackage{hyperref}
\usepackage{latexsym}
\usepackage{amsthm}
\usepackage{amssymb}

\DeclareGraphicsExtensions{.jpg,.pdf, .mps, .png, .eps, .ps, .EPS,.gif}

\begin{document}
\def\be{\begin{equation}}
\def\ee{\end{equation}}

\def\bc{\begin{center}}
\def\ec{\end{center}}
\def\bea{\begin{eqnarray}}
\def\eea{\end{eqnarray}}
\newcommand{\avg}[1]{\langle{#1}\rangle}
\newcommand{\Avg}[1]{\left\langle{#1}\right\rangle}

\def\ie{\textit{i.e.}}
\def\etal{\textit{et al.}}
\def\m{\vec{m}}
\def\G{\mathcal{G}}

\title{Percolation in  networks of networks \\
with random matching of nodes in different layers}
\author{ Ginestra Bianconi}

\affiliation{
School of Mathematical Sciences, Queen Mary University of London, London, E1 4NS, United Kingdom}
\author{Sergey N. Dorogovtsev}
\affiliation{Departamento de Fisica da Universidade de Aveiro, 13N, 3810-193, Aveiro, Portugal\\
A. F. Ioffe Physico-Technical Institute, 194021 St. Petersburg, Russia}

\begin{abstract}

We consider robustness and percolation properties of the networks of networks, in which random nodes in different individual networks (layers) can be interdependent. We explore the emergence of the giant mutually connected component, generalizing the percolation cluster in a single network to interdependent networks, and observe the strong effect of loops of interdependencies. In particular, we find that the giant mutual component does not emerge in a loop formed by any number of layers. In contrast, we observe multiple hybrid transitions in networks of networks formed by infinite number of randomly connected layers, corresponding to the percolation of layers with different number of interdependencies. In particular we find that layers with 
many 
interdependencies are more fragile than layers with 
less interdependencies. These hybrid transitions, combining a discontinuity and a singularity, are responsible for joining a finite fraction of nodes in different layers to the giant mutually connected component. In the case of partial interdependence, when only a fraction of interlinks between layers provide interdependence, some of these transitions can become continuous.  
\end{abstract}

\maketitle


\section{Introduction}

In order to unveil the full extent of the complexity of real social, technological, and biological systems, it is important to consider networks of networks, i.e. multilayer structures formed by interaction between different networks \cite{PhysReports,review2}.

Real-world networks \cite{Thurner,Mucha,Bullmore2009,Boccaletti} rarely live in isolation, and technological systems such as infrastructures, or biological systems such as a single cell or the brain cannot be fully understood if the multilayer perspective is not taken into account.  
Recent results \cite{Havlin1,Dorogovtsev,Son,Havlin2, HavlinEPL,Stanleyint,Goh,Kabashima,JSTAT,Gao1,Gao2,Gao3, Gao4,BD1,BD2,Weak,Kcore} on the robustness of multilayer networks pose new challenge for the design of better infrastructures where the effect of random damage and the risk of dramatic avalanches of failures can be reduced.
A pivotal role in this respect has been played by the observation that multilayer networks might be structurally more fragile than single layers and prone to dramatic avalanches of cascading failures \cite{Havlin1}.
It has been shown that the robustness of a multilayer network can be characterized by evaluating the size of the {\em mutually connected giant component} of these networks as a function of the probability of the initially inflicted damage to the network.
It has been found that on multilayer networks, critical phenomena \cite{crit} can show surprising new physics. In fact, differently from the giant component of a single network, the mutually connected component emerges  discontinuously at a hybrid transition, combining discontinuity and singularity, and close to this transition the system can be affected by global cascades o failures~\cite{Havlin1,Dorogovtsev}.
This result is not only interesting from the theoretical point of view, but it also raises some questions concerned to the robustness of multilayer biological networks, that would not have 
survived the evolutionary pressure if they 
were so fragile \cite{Makse,Liaisons}.
In Ref.~\cite{Makse}, for example, the case of multilayer brain network was considered and key structural correlation of the network have been detected 
responsible 
for restoring the robustness of biological networks. 

\begin{figure}
\begin{center}
$\begin{array}{c}
{\includegraphics[width=3.16in]{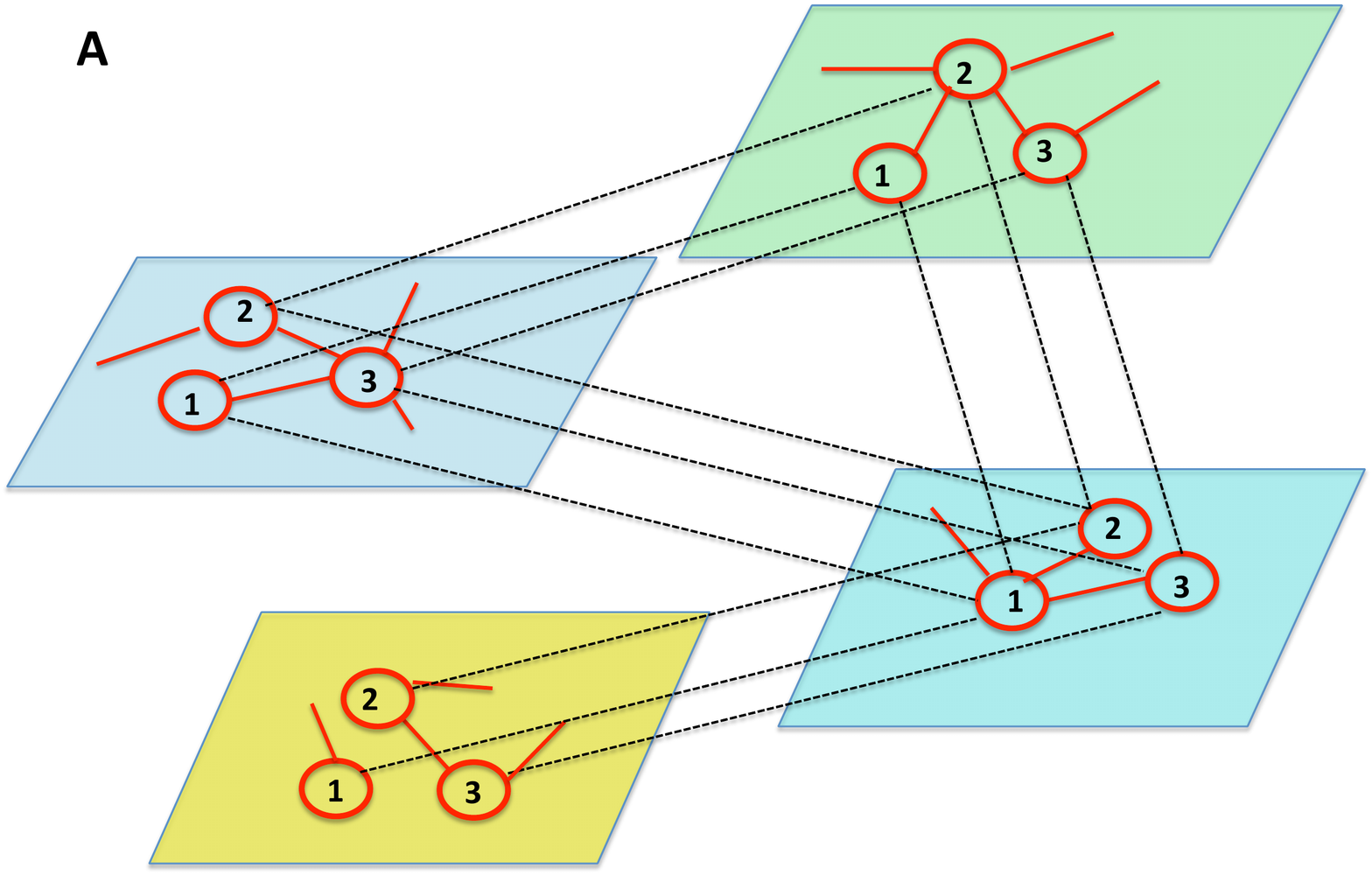}}\\
{\includegraphics[width=3.16in]{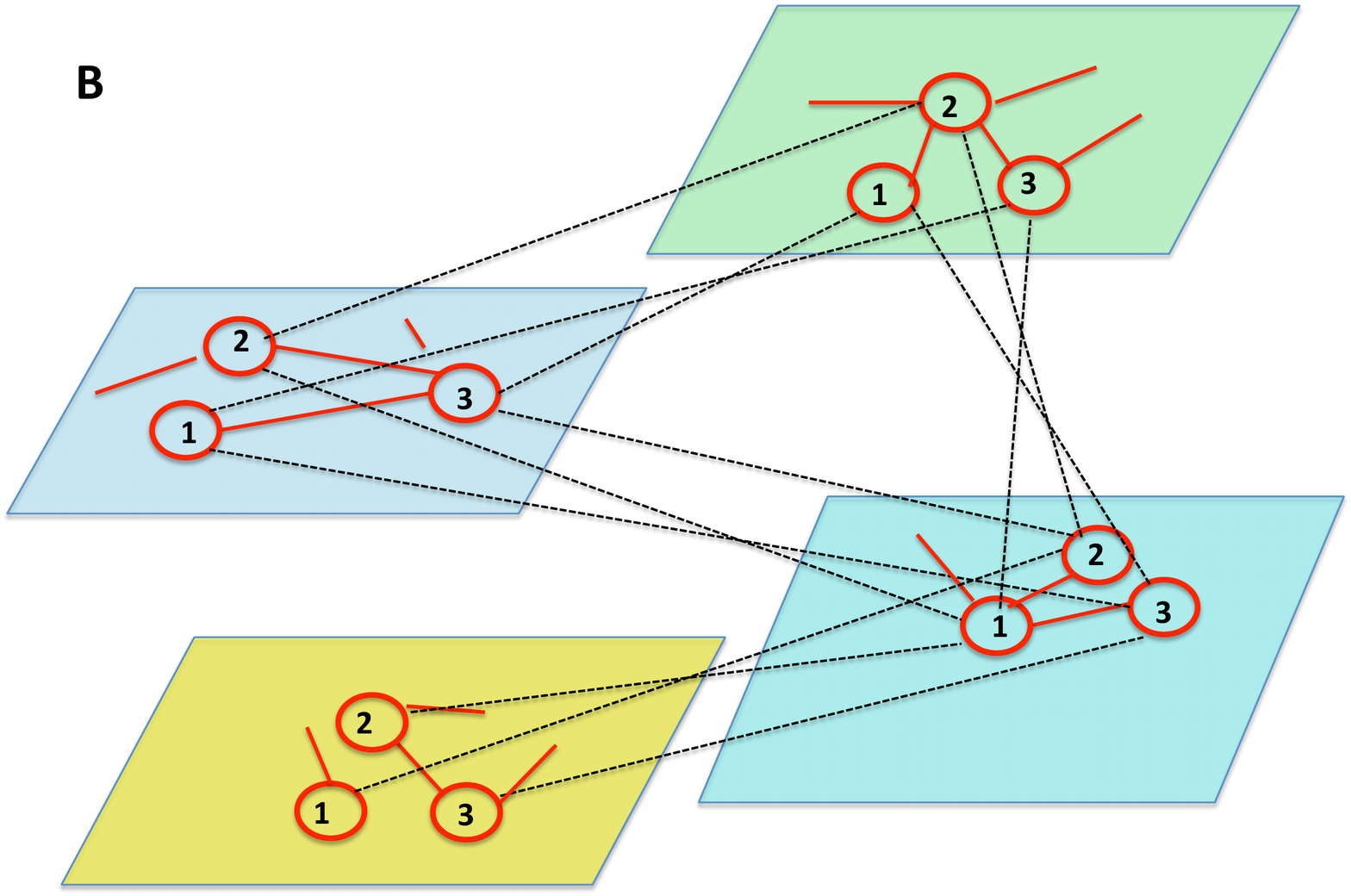}}\\
{\includegraphics[width=3.16in]{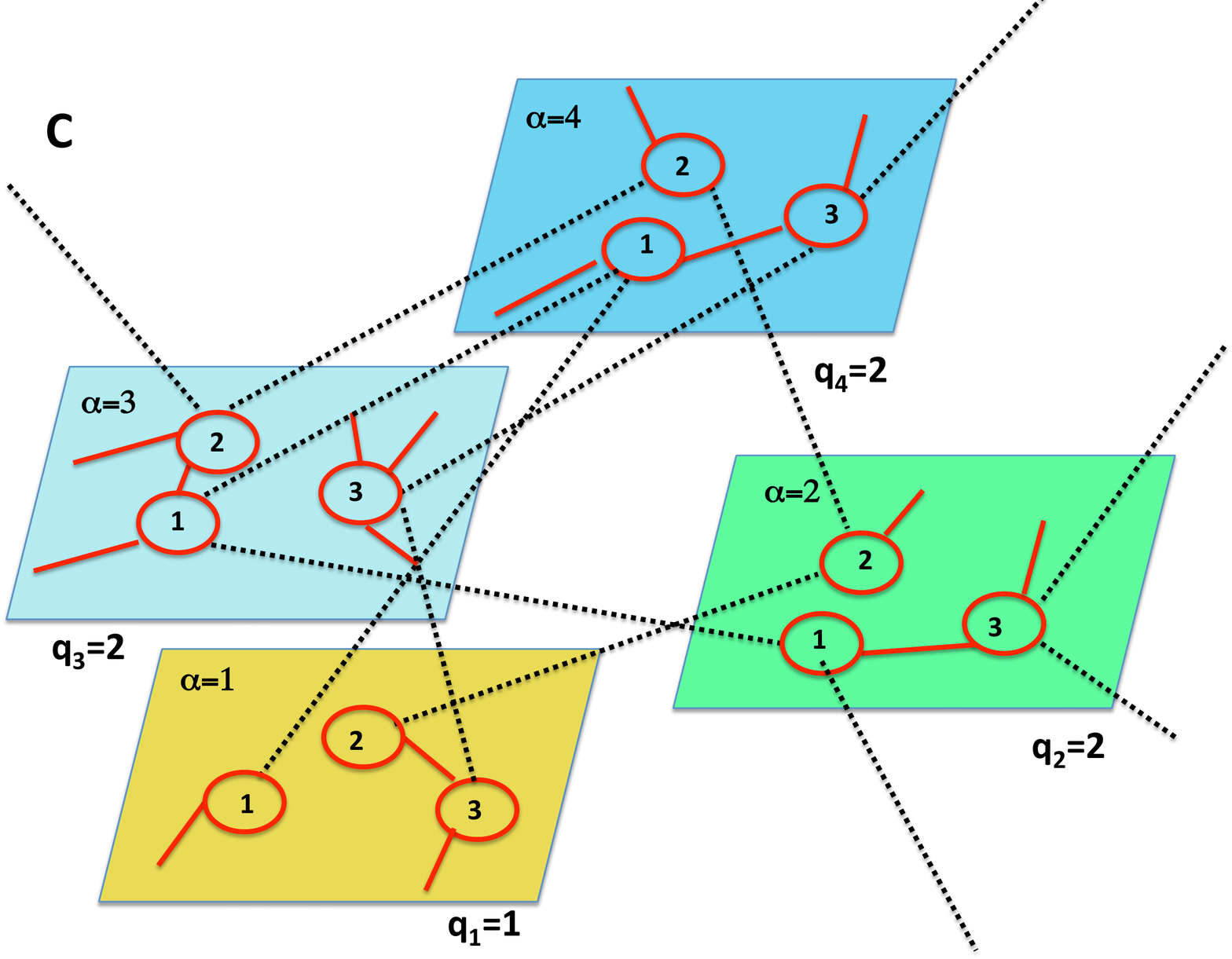}}
\end{array}$
\end{center}
\caption{(Color online)
Schematic view of various models of the networks of networks. In panel A the nodes in the network of networks  can be linked only to their replica nodes in other layers, see the work \cite{BD1}. In panel B the nodes in each layer of the network of networks can be linked to random nodes in an interdependent network (layer), this network of networks is considered in the present work. In panel C the nodes of each layer have fixed number of interlinks. The nodes of each layer can be linked only to replica nodes but these nodes can be chosen in any layer. This configuration model of a network of networks has been considered in Ref.~\cite{BD2}.}
\label{f1}
\end{figure}

While most of the results \cite{Havlin1,Dorogovtsev, Son,Havlin2,HavlinEPL,Stanleyint} concerning the emergence of the mutually connected component regard multiplex networks \cite{PRE}, i.e. networks in which the same set of nodes interact through different layers, here we consider the case of networks of networks \cite{Math,PhysReports,review2} in which the nodes in the different sub networks (layers) are different entities. For example, a network of networks can be formed by interacting power-grid and the Internet in which the nodes are power-plants and routers, or in biology a network of networks can be formed by interacting protein--protein interaction networks, transcription networks, or metabolic networks where the nodes 
represent different molecules.
Several results were already obtained for the robustness of networks of networks \cite{Gao1,Gao2,Gao3, Gao4,BD1,BD2}. In particular, in Ref.~\cite{Gao1} it has been shown that  
the emergence of the mutually connected component might depend on the presence 
of loops 
of interdependencies between layers. 
In Ref.~\cite{BD1} we focused on the networks of networks in which each node has its replica nodes in all other layers, and only the replica nodes can be interdependent, see Fig.~\ref{f1}A,C. In the present work we relax this constraint, which allowed interdependencies only between replica nodes, and study networks of networks in which there is no replica nodes, and random nodes in pairs of layers can be interdependent, see Fig.~\ref{f1}B. Clearly, in the absence of loops of interdependencies, the latter network of networks is readily reduced to the super tree from Ref.~\cite{BD1}, in which super nodes are different layers, with interdependent replica nodes. On the other hand, the networks of networks with random matching of nodes in different layers cannot be reduced to already considered models if these networks of networks contain loops of interdependencies. This is the case of our particular interest. 
Here we provide an exhaustive account of the robustness of networks of networks, stressing the role of interlinks (links between nodes of different layers, that is interdependencies) in determining the emergence of the mutually connected component.
In particular, we compare the 
model of a network of networks proposed by Gao et al. \cite{Gao1} with the case of a network of networks in which, in contrast to the work of Gao et al. \cite{Gao1}, each node can only link to a replica node in the other layer \cite{BD1}, and we show that the difference 
the ensembles of networks of networks dramatically changes the nature of the percolation transition. 

We confirm the result of Gao et al. \cite{Gao1} that the robustness of networks of networks 
without replica nodes depends on the loop structure of the super network (the network of interdependencies between the layers) while this phenomenon is not observed in the 
networks of networks 
with replica nodes. 
 
We also show that 
the mutually connected component can be subjected to multiple phase transitions instead of a single one.
In particular in the case of a random super network in which each layer has a given superdegree (number of interlinks of the nodes of a layer) there are multiple phase transitions corresponding to the activation/deactivation of layers of different superdegrees. At each of these transitions, a finite fraction of nodes in layers of specific superdegree join the giant mutually connected component. 
In particular we find that in this case layers with 
high superdegree $q$ are more fragile than layers with lower superdegree. 
This is quite opposite to what occurs 
in single networks where nodes of high degree are more robust in the sense that they are typically the last ones to be separated from the giant connected component by random damaging. 
This 
puzzle can be easily explained. 
Indeed, in a single network one node is active if at least one linked node is active, giving more robustness to high degree nodes. In contrast to this, in interdependent networks a node in one layer is active only if all the nodes interdependent on it  in the other layers are also active, 
making the nodes in the layers with high superdegree more fragile.

The multiple percolation phase transitions were previously observed in the configuration model of a network of networks \cite{BD2}, Fig.~\ref{f1}C, where, as in the present work, layers with many interdependencies turned out to be more fragile than layers with less interdependencies. 
One should notice that multiple phase transitions have been recently observed in percolation problems for networks with nontrivial clustering or with communities \cite{Bogunamultiple,Wumultiple}. 
In our case, the reason for the multiple phase transitions is specific percolation in a system of layers in a network of networks, which differs principally from these works. 

\section{Network of networks with random matching of nodes in interdependent 
layers}

Let us define the specific class of networks of networks that we study in this work, see Fig.~\ref{f1}B. 
The network of networks of this kind is formed by $M$ networks (layers) $\alpha=1,2\ldots, M$ each 
 of $N$  nodes $i=1,2\ldots, N$.  
 We assume that any layer $\alpha$ is linked to other layers $\beta$ 
 forming the super network ${\cal G}$ with the adjacency matrix ${\cal A}_{\alpha,\beta}$. 

 Every node $(i,\alpha)$ can be connected to a single node in 
 an interconnected network (layer) $\beta$ for which 
 ${\cal A}_{\alpha,\beta}=1$. If layer $\alpha$ is interdependent on layer $\beta$, then any  node $(i,\alpha)$ is connected with a node $(j,\beta)$ where $j=\pi_{\alpha,\beta}(i)$ and $\pi_{\alpha,\beta}$ is a random permutation of the labels $\{i\}$ of the  nodes. 
We 
assume that the 
interdependence is reciprocal, that is, if $j=\pi_{\alpha,\beta}(i)$ then $\pi_{\beta,\alpha}(j)=i$, i.e. $\pi_{\beta,\alpha}$ is the inverse permutation of $\pi_{\alpha,\beta}$.  
We define  the super-adjacency matrix of the network of networks as the matrix with elements $a_{i\alpha,j\beta}$, where $a_{i\alpha,j\beta}=1$ if there is a link between node $(i,\alpha)$ and node $(j,\beta)$ and zero otherwise \cite{BD1,BD2,Math}. 

The  mutually connected component is defined as follows. 
A node $(i,\alpha)$ is in the mutually connected component if the following conditions are met:
\begin{itemize}
\item[(a)]
this node has at least one neighbour $(j,\alpha)$ in its layer $\alpha$, which belongs to the mutually connected component;
\item[(b)]  
all the  
nodes $(\pi_{\alpha,\beta}(i),\beta)$, connected to node $(i,\alpha)$, in the interdependent networks $\beta$ for which ${\cal A}_{\alpha,\beta}=1$ are also in the mutually connected component.
\end{itemize}

Let us apply the cavity method (message passing equations approach) \cite{Kabashima,Mezard,Weigt,Son,BD1,BD2} to our problem. 
For a given network of networks, it is easy to construct a message passing algorithm indicating  if node $(i,\alpha)$ is in the mutually connected component. 
Let us denote by $\sigma_{i\alpha\to j\alpha}=1,0$ the  message 
within a layer, from node $(i,\alpha)$ to node $(j,\alpha)$, 
where $\sigma_{i\alpha\to j\alpha}=1$ 
provided node $(i,\alpha)$ is in the mutually connected component if we remove the link $(j,\alpha)$ in network $\alpha$. 

Furthermore, let us denote by $S'_{i\alpha\to \pi_{\alpha,\beta}(i),\beta}=0,1$ the  message 
that the node $(i,\alpha)$ sends to the interconnected with it node $(\pi_{\alpha,\beta}(i),\beta)$  in layer $\beta$. The message $S'_{i\alpha\to \pi_{\alpha,\beta}(i)\beta}=1$ indicates if the node $(i,\alpha)$ is in the mutually connected component when we remove the link between node $(i,\alpha)$ and node $(\pi_{\alpha,\beta}(i),\beta)$.
 In addition, we  assume that 
 node $(i,\alpha)$ can be removed or not removed from the network. We indicate with  $s_{i\alpha}=0$ a node 
 removed from the network, otherwise 
 $s_{i\alpha}=1$.
The message passing equations for these messages are given by 
\bea
\sigma_{i\alpha\to j\alpha}&=&s{i\alpha}\prod_{\beta\in {\cal N}(\alpha)}S'_{ \pi_{\alpha,\beta}(i) \beta \to i \alpha}
\nonumber 
\\
&&\times\left[1-\prod_{\ell\in Ne{\alpha}(i)\setminus j}(1-\sigma_{\ell\alpha\to i\alpha})\right]
,
\nonumber \\
S'_{i\alpha\to i\beta}&=&s{i\alpha}\prod_{\gamma\in {\cal N}(\alpha)\setminus \beta}S'_{\pi_{\alpha,\gamma}(i)\gamma \to i \alpha}
\nonumber 
\\
&&\times \left[1-\prod_{\ell\in Ne{\alpha}(i)}(1-\sigma_{\ell\alpha\to i\alpha})\right]
,
\label{mp1}
\eea
where $N_{\alpha}(i)$ indicates the set of nodes $(\ell,\alpha)$ which are neighbors of node $i$ in network $\alpha$, and ${\cal N}(\alpha)$ indicates the layers that are interdependent on network $\alpha$.
Finally $S_{i\alpha}$ indicates if a node $(i,\alpha)$ is in the mutually  
connected component ($S_{i\alpha}=1,0$). 
This indicator function can be expressed in terms of the messages as 
\bea
S_{i\alpha}&=&s_{i,\alpha}\prod_{\beta\in {\cal N}(\alpha)}S'_{\pi_{\alpha,\beta}(i)\beta \to i\alpha} 
\nonumber 
\\[5pt]
&&\times \left[1-\prod_{\ell\in N_{\alpha}(i)}(1-\sigma_{\ell\alpha \to i\alpha})\right]
.
\label{S}
\eea 
The solution of these equations implies that a node $(i,\alpha)$ is in the mutually connected giant component if and only if 
\begin{itemize}

\item[(a)] 
it has at least one neighbor in layer $\alpha$ that is in the mutually connected component;
\item[(b)] 
all the nodes $(j,\gamma)$  
that can be reached 
through a chain of interdependence links starting from node $(i,\alpha)$ 
have at least one neighbor in their layer $\gamma$ 
sitting in the mutually connected component. In the following we 
indicate by ${\cal C}(i,\alpha)$ the set of nodes $(j,\gamma)$ that can be reached from node $(i,\alpha)$ by 
interdependence links.
\end{itemize}
This result can be 
given mathematically by  
expressing the interlayer message $\sigma_{i\alpha\to j\alpha}$ in terms of the other interlayer messages, namely 
\bea
\hspace{-10mm}
\ \ \ \ \ \ 
\sigma_{i\alpha\to j\alpha}&=&s_{i\alpha}\!\!\!\!\prod_{(i^{\prime},\gamma)\in {\cal C}(i,\alpha)}\!\!\left\{\!s_{i^{\prime}\gamma}\!\!\left[1-\!\!\!\prod_{\ell\in N_{\gamma}(i^{\prime})}\!\!(1{-}\sigma_{\ell\gamma\to i^{\prime}\gamma})\!\right]\!\!\right\}
\nonumber \\[5pt]
&&\times\left[1-\prod_{\ell\in N_{\alpha}(i)\setminus j}(1-\sigma_{\ell\alpha\to i\alpha})\right].
\label{mp2}
\eea
Finally the probability $S_{i\alpha}$ that a node is in the mutually connected component can also be expressed only in terms of the interlayer messages giving 
\bea
\hspace*{-10mm}
S_{i\alpha}&=&s_{i\alpha}\!\!\!\!\prod_{(i^{\prime},\gamma)\in {\cal C}(i,\alpha)}\!\!\left\{s_{i^{\prime}\gamma}\!\!\left[1-\!\!\prod_{\ell\in N_{\gamma}(i^{\prime})}\!\!(1{-}\sigma_{\ell\gamma\to i^{\prime}\gamma})\right]\!\!\right\}
\nonumber \\[5pt]
&&\times\left[1-\prod_{\ell\in N_{\alpha}(i)}(1-\sigma_{\ell\alpha\to i\alpha})\right].
\label{S2}
\eea

\section{Difference between a tree supernetwork and a supernetwork with loops}

For the specific  
topology of the network of networks formed by $M$ layers, that we consider, there is a clear difference  between a tree supernetwork and a supernetwork with loops as observed also in \cite{Gao1}. 
For a tree supernetwork the connected components of interdependent nodes ${\cal C}(i,\alpha)$ have all equal size, they are formed by a single  node in each layer of the supernetwork, and their cardinality is $|{\cal C}(i,\alpha)|=M$. Here we assume that the supernetwork consists of a single component. 
Instead, for a network of networks 
containing loops the cardinality of the connected components of interdependent nodes $|{\cal C}(i,\alpha)|$ can 
differ significantly 
from $M$ and can  
fluctuate from node to node. In fact ${\cal C}(i,\alpha)$ can contain several nodes in each layer $\alpha$. 
As a simple example one can indicate a multiplex network of $M=3$ in which each node in a layer is interdependent on its replica nodes in each of the other $M-1=2$ layers. This configuration is based on a supernetwork of three layers formed by a loop of size $M=3$. 

In this multiplex network 
percolation turns out to be  
very different from 
the network of networks with random interconnections based on the same supernetwork (loop of $M=3$), see the subsequent section. 
The reason for this difference is just that 
in the multiplex network the nodes can only link to their replica nodes in contrast to the networks of networks with random interconnections.


\section{Tree supernetwork}\label{iv}


\subsection{The simplest model}\label{iva}

Let us 
assume that each layer $\alpha$ is formed by random networks with a degree distribution $P_{\alpha}(k)$.
Moreover  
we assume that the initial damage inflicted to the network is random, and that the probability $P(\{s_{i\alpha}\})$ is given by 
\bea
P(\{s_{i\alpha}\})=\prod_{\alpha}\prod_i p^{s_{i\alpha}}(1-p)^{1-s_{i\alpha}}.
\label{Ps1}
\eea
Averaging the message passing equations over the ensemble of networks with a given tree supernetwork, we 
find that the probability $\sigma_{\alpha}=\Avg{\sigma_{i\alpha \to j\alpha}}$ that by following a link in layer $\alpha$ we reach a node in the connected component is given by  
\bea
\sigma_{\alpha}=p^M\prod_{\beta\neq \alpha}[1-G_0^{\beta}(1-\sigma_{\beta})][1-G_1^{\alpha}(1-\sigma_{\alpha})],
\label{t1}
\eea
where we have indicated by $G_0^{\alpha}(x)$ and $G_1^{\alpha}(x)$, respectively, the generating functions 
\bea
G_0^{\alpha}(x)&=&\sum_k P_{\alpha}(k)x^k,
\nonumber 
\\[5pt]
G_1^{\alpha}(x)&=&\sum_k \frac{k}{\avg{k}_{\alpha}}P_{\alpha}(k) x^{k-1},
\eea
where $\avg{k}_{\alpha}$ is the average degree of layer $\alpha$.
Furthermore, the probability that a node in layer $\alpha$ is in the mutually connected component is given by 
\bea
S_{\alpha}=p^M\prod_{\beta}[1-G_0^{\beta}(1-\sigma_{\beta})].
\label{t2}
\eea
Equations (\ref{t1})--(\ref{t2}) are the same that determine the mutually connected component in a network of networks in which any node in a layer can be interdependent only on its replica nodes in 
another layer and which is based. 
The reason for this coincidence was explained in the introduction. 
In particular, these equations are the same as 
for a multiplex network, i.e. for a network of networks with interdependent replica nodes, based on a fully connected supernetwork. 
Finally, if all the 
layers have the same degree distribution 
and so the generating functions $G_0^{\alpha}(x)=G_0(x) \ \forall \alpha$, $G_1^{\alpha}(x)=G_1(x) \ \forall \alpha$, then the average message and the order parameters are the same for every layer, i.e. $\sigma=\sigma_{\alpha}\ \forall \alpha$ and $S=S_{\alpha}\ \forall \alpha$, and they are given by 
\bea
\sigma&=&p^M[1-G_0(1-\sigma)]^{M-1}[1-G_1(1-\sigma)],
\nonumber 
\\[5pt]
S&=&p^M[1-G_0(1-\sigma)]^{M}.
\label{t3}
\eea


\subsection{Tree supernetwork with partial interdependencies of the nodes}\label{ivb}

It is interesting to study the role of partial interdependencies on the robustness of network of networks with tress supernetwork.
The concept of partial interdependencies has been introduced and discussed already in \cite{Havlin2,HavlinEPL,Stanleyint,Son,Gao1,BD2}.
If each node is interdependent on  
a node linked to 
it in the other 
layer with probability $r$ (equivalently one can say that the fraction $1-r$ of the interlinks is removed), then Eqs.~(\ref{t1}), (\ref{t2}) become, respectively, 
\bea
\sigma_{\alpha}=p^M\prod_{\beta\neq \alpha}[1-rG_0^{\beta}(1-\sigma_{\beta})][1-G_1^{\alpha}(1-\sigma_{\alpha})]
,
\nonumber 
\\[5pt]
S_{\alpha}=p\prod_{\beta\neq \alpha}[1-rG_0^{\beta}(1-\sigma_{\beta})][1-G_0^{\alpha}(1-\sigma_{\alpha})].
\label{t4}
\eea
If  
all the layers have the same degree distribution, then we obtain, instead of Eq. (\ref{t3}),  
\bea
\sigma&=&p^M[1-rG_0(1-\sigma)]^{M-1}[1-G_1(1-\sigma)],
\nonumber 
\\[5pt]
S&=&p^M[1-rG_0(1-\sigma)]^{M-1}[1-G_0(1-\sigma)].
\eea
While at $r=1$, the percolation transition in these networks is hybrid, combining a discontinuity and singularity, due to the partial interdependence 
a tricritical point $(p_c,r_c)$ takes place, where discontinuity disappears.  
This point is obtained from the following equations:
\bea
\qquad
&&\hspace*{-12mm}1-\left.\frac{d}{d\sigma}\left\{p^M[1{-}rG_0(1{-}\sigma)]^{M-1}[1{-}G_1(1{-}\sigma)]\right\}\right|_{\sigma=0}\!\!\!\!=0
,
\nonumber 
\\[5pt]
&&\hspace*{-12mm}\left.\frac{d^2}{d\sigma^2}\left\{p^M[1{-}rG_0(1{-}\sigma)]^{M-1}[1{-}G_1(1{-}\sigma)]\right\}\right|_{\sigma=0}\!\!\!\!=0,
\eea 
and for 
$r\leq r_c$  
the percolation transition  
is continuous.


\subsection{Properties of the percolation transition in a tree supernetwork}\label{ivc}

Based on Secs.~\ref{iva} and \ref{ivb}, we conclude that 
the percolation transition in a network of networks in the case of a tree supernetwork has the following properties:

\begin{itemize}
\item The size of the mutually connected giant component depends only on the number of layers $M$ but not on the other details of the structure of the tree supernetwork.
\item
If the degree distribution of each layer is the same, all the networks either contain a giant cluster or none does.
\item
The mutually connected components appears discontinuously if there is no partial interdependence.
\item
In the presence of a probability $r$ that an interlink indicates 
an actual interdependence, there is a critical value $r=r_c$ such that for $r<r_c$ the percolation transition becomes continuous.
\item 
The equations determining the size of the mutually connected component are the same as the ones valid for a  network of networks in which the nodes can be linked only to their replica nodes in other layers, as, e.g., in the case of a multiplex network.
\end{itemize}


\section{Loop supernetwork}


\subsection{The simplest case}

The case of a supernetwork with loops 
significantly differ from a tree supernetwork since the size of the connected components of interdependent nodes can be actually much larger than $M$ and 
fluctuates from node to node. 
Let us consider for simplicity a loop supernetwork formed by $M$ layers of $N$ nodes in each layer.  The loops is labelled in such way that layer $\alpha$ is connected with layer $\alpha+1$ and layer $\alpha-1$, where we identify layer $\alpha=0$ with layer $\alpha=M$ and layer $\alpha=-1$ with layer $\alpha=M-1$.
Therefore we assume that each node in layer $\alpha$ is connected with a node in layer $\alpha+1$ and a node in layer $\alpha-1$, but that these nodes are chosen randomly.
For this network of networks, a node $(i,\alpha)$ is in the mutually connected component if and only if at least one its neighbor within the same layer is in the mutually connected component and if all the nodes that can be reached by interdependencies links from this node [their set is denoted by ${\cal C}(i,\alpha)$] are in the mutually connected component. Note that the size of ${\cal C}(i,\alpha)$ is not determined by $M$ since a connected component of the interlinks can be larger than $M$. 
Clearly, 
the connected component ${\cal C}(i,\alpha)$ of any node $(i,\alpha)$ must be a loop in these networks of networks. 
Let us calculate the probability that  
this loop has size $s$.
We start from a node $(i,\alpha)$ in layer $\alpha=1$,
and 
proceed in one direction of the loop after $M$ steps (i.e., after one complete turnover) we come back to layer $\alpha=1$. Since the 
supernetwork is a loop and since each node that we are visiting in this path has ``superdegree'' 2 (number of its interlinks) we cannot visit these nodes more than once in the path. Therefore the only possibility 
to close a loop is that, moving all the time forward, we will finally return to  
the starting node. The probability that we do not meet the initial node after one round is 

\bea 
(1-1/N).
\eea

The probability that we do not meet this node at the  $s$-th round is 
\bea
[1-1/(N-(s-1))],
\eea
because at each round we can only connect either to the original node or to one of the not visited nodes of the network. 

Proceeding in this way we 
find that the probability that 
starting from layer $\alpha=1$ we end in this layer 
after $s$ rounds along the loop is given by the expression 
\bea
P(s)=\left\{\begin{array}{lcc}\prod_{i=0}^{s-2}\left(1-\frac{1}{N-i}\right)\frac{1}{N-(s-1)}&\mbox{for}& s>1
\nonumber 
\\[8pt]
\frac{1}{N} & \mbox{for}& s=1\end{array}\right.
\label{psl}
\eea
Simplifying this expression it is easy to find that $P(s)$ is uniform, i.e. 
\bea
P(s)=\frac{1}{N}.
\eea
We  
assume that every layer has the same degree distribution, 
so averaging over the ensemble we 
obtain 
the following equation for $\sigma_{\alpha}=\sigma$:  
\bea
\sigma=p\sum_{s=1}^{N} P(s)[1-G_0(1-\sigma)]^{Ms-1} [1-G_1(1-\sigma)]. 
\eea
Performing the sum, we obtain 
\bea
\sigma&=&\frac{p}{N}\frac{1-[1-G_0(1-\sigma)]^{M(N+1)}}{1-[1-G_0(1-\sigma)]^M}
\nonumber 
\\[5pt]
&&\times[1-G_0(1-\sigma)]^{M-1}[1-G_1(1-\sigma)]
, 
\eea
which leads to $\sigma=0$.
Therefore in this networks for $N\to \infty$ we never observe the mutually connected component.
This can be also understood intuitively by making the following observation. 
In any 
layer the probability that by following a uniformly randomly chosen link we reach a node $(i,\alpha)$ belonging to a component  ${\cal C}(i,\alpha)$ of cardinality $|{\cal C}(i,\alpha)|=sM$ is $P(s)$. 
Since $P(s)$ is uniform, i.e. $P(s)=1/N $ with $s\in[1,N]$,  
the probability that a component ${\cal C}(i,\alpha)$ is formed by a finite number of layers is vanishing as $N\to \infty$. This means that the clusters formed by interlinks are typically diverging with the system size $N$ implying the absence of the mutually connected component in these systems.


\subsection{Loop with partial interdependence}

We showed above that 
in the absence of partial interdependence, the probability that a randomly chosen node belongs to a loop of size $sM$ is $P(s)=1/N$.
Given the partial interdependencies, one can be in three situations, namely either the node belongs to a loop  formed by interdependency links that are not removed, or the node belongs to a loop with only one link removed, or the node belongs to a segment of a loop starting and ending on two points having partial interdependencies.
In the first case the probability $P(S)^{(1)}$ that a random node  $(i,\alpha)$ belongs to a connected component of interdependency links of size $S=|{\cal C}(i,\alpha)|$ is given by
\bea
\hat{P}^{(1)}(S|s)=\frac{1}{N}r^{S}\delta(S,Ms),
\eea
with $s=1,2\ldots, N$.
In the second case  the node belongs to a loop of size $s$ where only one link has been removed due to the partial interdependence.  
The probability that a random node $(i,\alpha)$ belongs to a connected component of this type is given by 
\bea
\hat{P}^{(2)}(S|s)=\frac{S}{N}r^{S-1}(1-r)\delta(S,Ms)
,
\eea
where we have taken into account the fact that the single removed link 
can be any of the $Ms$ interlinks forming the loop.
Finally in the case in which the random node belongs to a segment of the loop of size $Ms$ we have 
\bea
\hat{P}^{(3)}(S|s)=\frac{1}{N}S(1-r)^2r^{S-1}\theta(Ms-S)
,
\eea
where $\theta(x)=x$ if $x\leq 0$ and $\theta(x)=1$ if $x>0$.
Therefore the probability that a random node belongs to a connected component of size $S$ is given by 
\bea
\hat{P}(S)=\sum_{s=1}^N\left[\hat{P}^{(1)}(S|s)+\hat{P}^{(2)}(S|s)+\hat{P}^{(3)}(S|s)\right].
\eea
Finally, averaging over the ensemble, and considering these  different possibilities, we have 
\bea
\!\!\!\!\!\!\!\!\!\!\!\!\!\!
&&\sigma=p\sum_{S=1}^{NM}\hat{P}(S)[1-G_0(1-\sigma)]^{S-1} [1-G_1(1-\sigma)].
\label{rl}
\eea

Let us define the function
\bea
\!\!
h(x)=x-p\sum_{S=1}^{NM}\hat{P}(S)[1-G_0(1{-}x)]^{S-1}[1-G_1(x)].
\eea
For $r=0$ the mutually connected component reduces to the giant component of a single layer having an ordinary continuous percolation transition. Therefore we assume that there is a critical value of $r$, i.e. $r=r_c$ such that for all $r<r_c$ the model has a second order phase transition.
By imposing the condition $h^{\prime}(0)=0$, we determine the position of these phase transitions, while the tricritical point $r=r_c$ can be found by imposing  $h^{\prime}(0)=h^{\prime\prime}(0)=0$.
We found therefore that the second order critical point occurs at 
\bea
p\frac{\Avg{k(k-1)}}{\avg{k}}\hat{P}(1)=1,
\eea
and inserting the value of $\hat{P}(1)$ we get
\bea
p\frac{\Avg{k(k-1)}}{\avg{k}}(1-r)^2=1.
\eea
Note that for $r=0$ this is the well known percolation threshold of single networks.
At the tricritical point  we also have $h^{\prime\prime}(0)=0$ giving the condition
\bea
2\hat{P}(2)\Avg{k(k-1)}\avg{k}=\hat{P}(1)\Avg{k(k-1)(k-2)}.
\eea
Inserting the values of $\hat{P}(1)$ and $\hat{P}(2)$ we get 
\bea
4(1-r)^2r\Avg{k(k-1)}\Avg{k}=(1-r)^2\Avg{k(k-1)(k-2)}.
\eea
Therefore we found that 
$r_c$ is given by the following relation: 
\bea
r_c=\min\left(\!1,\,\frac{\Avg{k(k-1)(k-2)}}{4\Avg{k(k-1)}\Avg{k}}\right).
\eea
If every layer has a Poisson degree distribution with average degree $c$ we conclude 
that the transition is second-order for every $r<r_c=\frac{1}{4}$ with a critical point at 
\bea
p=\frac{1}{c}\,\frac{1}{(1-r)^2}.
\eea
On the contrary, for $r>r_c$, the transition is discontinuous.


\subsection{Properties of the percolation transition in a loop network}

We  
summarize the results obtained for a loop supernetwork formed by layers  
with the same number of nodes and the same degree distribution in the following points: 
\begin{itemize}
\item
If the supernetwork is a loop, and the interdependencies are random, the size of the mutually connected component is zero for any value of $M$, i.e. different from the size of the connected component of a   multiplex network formed by the same layers.
\item
The percolation transition can become continuous in the case of 
partial interdependence.
\item
In the presence of partial interdependence the discontinuous transition might dependent on $M$, the continuous transition is always independent on the value of $M$.
\item
All the layers percolate for the same value of $p$. In fact  either all the layers have a finite fraction of nodes in the mutually connected component or none of them has a finite fraction of nodes in the mutually connected component.
\end{itemize}


\section{Random supernetwork}


\subsection{The simplest case}

 We assume here that   
 each layer (network) $\alpha$ is generated from a configuration model with the same degree distribution $P_{\alpha}(k)=P(k)$, and that the supernetwork is a random network with given degree distribution $P(q)$. 
In particular we consider a superdegree sequence $\{q_{\alpha}\}$ drawn from the distribution $P(q)$ and 
$M$ degree sequences $\{k_i^{\alpha}\}$ with $i=1,\ldots, N$ and $\alpha=1,2,\ldots, M$ and construct a network of networks with a super-adjacency matrix ${\bf a}$ and  a supernetwork ${\cal A}$ with a probability $P({\bf a},{\cal A})$  given by 
\bea
&&\hspace*{-12pt}P({\bf a},{\cal A})\!=\!\frac{1}{Z}\!
\prod_{\alpha=1}^M\!\left\{\prod_{i=1}^N
\!\!\delta\!\!\left(\!k^{\alpha}_i,\!\sum_{j}a_{i\alpha,j\alpha}\!\right)
\!\!\delta\!\left({\cal A}_{\alpha,\beta},a_{i\alpha\pi_{\alpha,\beta}(i)\beta}\right)\right.
\nonumber 
\\[5pt]
&&
~~~~~~\times
\left.\delta\left(\sum_\beta {\cal A}_{\alpha,\beta},q_{\alpha}\right)\!\!\right\}
,
\label{Pa}
\eea
where $\delta(a,b)$ 
denotes the Kronecker 
symbol, 
$Z$ is the normalization constant, 
$\pi_{\alpha,\beta}(i)$ are random permutations  
provided that   $\pi_{\beta,\alpha}$ is the inverse permutation of $\pi_{\alpha,\beta}$.
Furthermore we assume that  the nodes  $(i,\alpha)$ are removed with probability $1-p$, i.e. we consider  the following expression for the probability $P(\{s_{(i,\alpha)}\})$ of the variables $s_{i\alpha}$:    
\bea
P(\{s_{i\alpha}\})=\prod_{\alpha}\prod_i p^{s_{i\alpha}}(1-p)^{1-s_{i\alpha}}
.
\label{Ps}
\eea
In order to quantify the expected  size of the mutually connected component in this ensemble, we 

average the messages over this ensemble of the network of networks.
The message-passing equations for this problems are  
Eqs.~(\ref{mp2}) that we rewrite here for convenience, 

\bea
&&\!\!\!\!\!\!
\sigma_{i\alpha\to j\alpha}=\!\!\prod_{(i^{\prime}\gamma)\in {\cal C}({i,\alpha})\setminus \alpha}\!\left\{s_{i^{\prime}\gamma}\left[1-\prod_{\ell\in N_{\gamma}(i)}(1-\sigma_{\ell\gamma\to i^{\prime}\gamma})\right]\right\}
\nonumber 
\\[5pt]
&&~~~~~~~\times\, s_{i\alpha}\left[1-\prod_{\ell\in N_{\alpha}(i)\setminus j}(1-\sigma_{\ell\alpha\to i\alpha})\right]
,
\nonumber 
\eea
where ${\cal C}(i,\alpha)$ is the connected component of the nodes that can be reached 
from node  $(i,\alpha)$ by interdependencies.
Therefore the equations for the average message within a layer are given in terms of the parameter $p=\Avg{s_{i\alpha}}$ and the generating functions $G_0^{k}(z)$, $G_1^{k}(z)$, $G_0^q(z)$, and $G_1^q(z)$ are given by
\bea
\!\!\!\!\!\! G_0^k(z)=\sum_k P(k) z^k,\  && G_1^k(z)=\sum_k \frac{kP(k)}{\Avg{k}}z^{k-1} 
,
\nonumber 
\\[5pt]
\!\!\!\!\!\! G_0^q(z)=\sum_q P(q) z^q,\  && G_1^q(z)=\sum_q \frac{qP(q)}{\Avg{q}}z^{q-1}.
\eea
In particular, if we indicate by $\sigma_q$ the average messages in a layer $\alpha$ of degree $q_{\alpha}=q$ we obtain 
\bea
&&\sigma_{q}=p\sum_{s}P(s|q)\left[\sum_{q'}\frac{q' P(q')}{\Avg{q}}p[1-G_0(1-\sigma_{q'})]\right]^{s-1}
\nonumber 
\\[5pt]
&&~~~\,\times[1-G_1(1-\sigma_{q})]
, 
\label{sq}
\eea
where $P(s|q)$ indicates the probability that a node $i$ in layer $\alpha$ with $q_{\alpha}=q$ is in a connected component ${\cal C}(i,\alpha)$ 
of cardinality  $|{\cal C}(i,\alpha)|=s$.
Similarly, the probability that a node $i$ in a layer $\alpha$ with super-degree $q_{\alpha}=q$ is in the mutually connected component, $S_q=\Avg{S_{i\alpha}}$, is given by 
\bea
&&S_q=p\sum_{s}P(s|q)\left[\sum_{q'}\frac{q' P(q')}{\Avg{q}}p[1-G_0(1-\sigma_{q'})]\right]^{s-1}
\nonumber 
\\[5pt]
&&~~~\,\times[1-G_0(1-\sigma_{q})]
.
\label{Sq}
\eea
Equations~(\ref{sq}) and (\ref{Sq}) are valid for any network of networks ensemble described by Eqs.~(\ref{Pa})--(\ref{Ps}). In the following we study, in particular, the case of the 
infinite number of layers $M$, 
in which the  supernetwork is locally tree-like.  
We take into account that the interlinks connect random nodes in the different layers, and the supernetwork is a random network formed by a number of layers going to infinity. Therefore the connected networks ${\cal C}(i,\alpha)$ are random networks with degree distribution $P(q)$. Note that in the configuration model of a network of networks with a number of layers $M\to \infty$,  considered in Ref.~\cite{BD2}, in which each layer has fixed superdegree 
but each node can be interdependent on any of its replica nodes in the other layers,  we also found that the connected networks ${\cal C}(i,\alpha)$ are random networks with degree distribution $P(q)$.
Then the distribution of the sizes of the connected interdependent networks ${\cal C}(i,\alpha)$ is the same as the one found for the configuration model of network of networks 
of Ref.~\cite{BD2}. 
Consequently this problem has the same solution of the configuration model of network of networks  
in which each node has a single replica nodes in each layer but the interlinks are random, see Fig.~\ref{f1}C. 

Therefore it can be shown, following the same derivation as in Ref.~\cite{BD2} that the equations characterizing the emergence of the giant mutually connected component are given by 
\bea
&&\sigma_q=p(\Sigma)^q[1-G_1^k(1-\sigma_q)] 
,
\nonumber
\\[5pt]
&&S_q=p(\Sigma)^q[1-G_0^k(1-\sigma_q)],
\nonumber
\\[5pt]
&&~\Sigma =\left[\sum_{q'}\frac{q' P(q')}{\Avg{q}}p[1-G_0^k(1-\sigma_{q'})]\right]
\nonumber 
\\[5pt]
&&~~~~\times \sum_{q'}\frac{q' P(q')}{\Avg{q}}(\Sigma)^{q'-1},
\eea
where $\Sigma$ is the order parameter of the transition. 
If $\Sigma=0$ then $\sigma_q=0$ $\forall q$, and none of the layers percolates.
Let us consider for simplicity 
a situation in which each layer is formed by a Poisson network with average degree $\avg{k}=c$. Then the previous equations become
\bea
&&\!\!\!\!\!\!\!\!\!\!\!\!\!\!\!\!\!\!\sigma_q=S_q=p(\Sigma)^q(1-e^{-c\sigma_q}) 
,
\nonumber
\\[5pt]
&&\!\!\!\!\!\!\!\!\!\!\!\!\!\!\!\!\!\!\Sigma=\left[\sum_{q'}\frac{q' P(q')}{\Avg{q}}p[1-e^{-c\sigma_{q'}}]\right]\sum_{q'}\frac{q' P(q')}{\Avg{q}}(\Sigma)^{q'-1}
.
\label{ps}
\eea
In the case of a regular supenetwork, i.e. $P(q)=\delta(q,m)$, we have $\sigma=\sigma_m=\sqrt{\Sigma}$  satisfying the equation
\bea
\sigma=p\sigma^{q/2}(1-e^{-c\sigma})
.
\eea
In the more general case of an arbitrary distribution $P(q)$ the problem defined in Eqs.~(\ref{ps}) continues to have  a single order parameter 
$\Sigma$. 
The first equation in Eqs.~(\ref{ps}) has solution expressed in terms of the principal value of the Lambert function $W(x)$, which is given by 
\bea
\sigma_q=\frac{1}{c}\left[pc (\Sigma)^q+W\left(-pc(\Sigma)^{q}e^{-pc (\Sigma)^{q}}\right)\right]
. 
\label{sigmaq}
\eea
Inserting this solution back into the equation for $\Sigma$ in Eqs.~(\ref{ps})  
we find
\bea
\!\!\!\!\!\!\!\!\!\!\!\!\!\!
&&\Sigma=\left[p +\sum_{q}\frac{q P(q)}{\Avg{q}}\frac{1}{c}(\Sigma)^{-q}W\left(-pc(\Sigma)^{q}e^{-pc (\Sigma)^{q}}\right)\right] 
\nonumber 
\\[5pt]
\!\!\!\!\!\!\!\!\!\!\!\!\!
&&~~~~~\times 
G_1^q(\Sigma)
.
\label{op}
\eea
\begin{figure}
\includegraphics[width=4.5in]{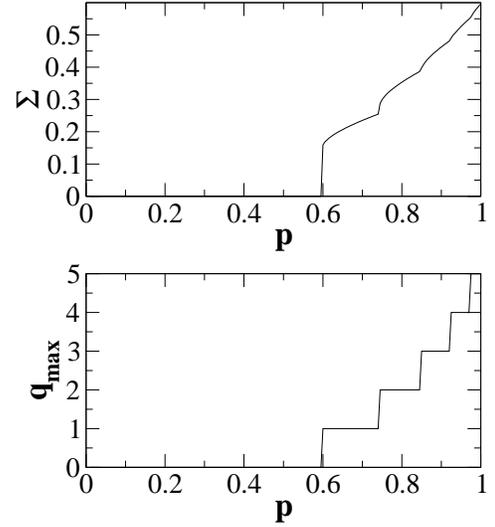}
\caption{(Color online) Order parameter $\Sigma$ of the multiple phase transition of the model, for a scale-free supernetwork with the power-law degree distribution exponent $\gamma=2.9$, in which each layer of the supernetwork is formed by a Poisson network with average degree $c=20$. In correspondence of each of the transitions, layers with higher superdegree $q$ are activated/deactivated. The figure also shows the value $q_{max}$ of the maximal superdegree of layers with a mutually connected giant component as a function of $p$, the probability that any node is not damaged by the initial perturbation. }
\label{f2}
\end{figure}
\begin{figure}
\begin{center}
\includegraphics[width=3.1in]{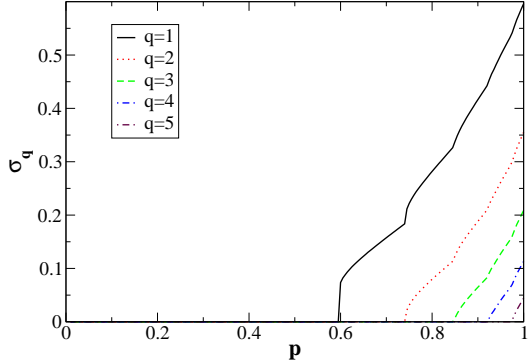}
\end{center}
\caption{(Color online) Probabilities $\sigma_q$ that by following a link in a layer with superdegree $q$ we reach a node in the giant mutual component as a function of the parameter $p$ indicating the probability that a node of the network is not damaged originally. 
The curves are shown for a scale-free supernetwork with the power-law degree distribution exponent $\gamma=2.9$, in which each layer formed by a Poisson network with average degree $c=20$. 
For this network of networks, $\sigma_q$ coincides with $S_q$, which is the fraction of nodes of a layer with superdegree $q$, belonging to the giant mutually connected component. 
For different transitions,  
layers with different superdegree are damaged in the network.}
\label{f3}
\end{figure}
This equation can be written as $F(\Sigma,p)=0$. By imposing the conditions $F(\Sigma,p)=0$ and $dF(\Sigma,p)/d\Sigma=0$ we can find the set of critical points $p=p_c$ at each of which there are 
discontinuities (jumps) of the order parameter $\Sigma(p)$, see Fig.~\ref{f2}. (Note that for the sake of brevity, hereafter we omit the index distinguishing these transitions from each other.) 
The  equation $dF(\Sigma,p)/d\Sigma=0$ reads as 
\bea
&&\hspace*{-11mm}1/p=\sum_{q}\frac{q P(q)}{\Avg{q}}(1-e^{-c\sigma_q})\sum_q \frac{q(q-1)P(q)}{\avg{q}}\Sigma^{q-2}
\nonumber 
\\[5pt]
 &&\hspace*{-5mm}+cpG_1^q(\Sigma)\!\!\sum_{q<q_{max}} \frac{q^2P(q)}{\avg{q}}e^{-c\sigma_q}\frac{\Sigma^{q-1}(1-e^{-c\sigma_q})}{1-pc\Sigma^qe^{-c\sigma_q}}
 ,
\eea
with $q_{max}=[-\log(pc)/\log(\Sigma)]$.
Analysing Eq.~(\ref{op}) for the order parameter we can show that, in addition to a jump, the order parameter $\Sigma$ has a singularity at each of the critical points $p=p_c,\Sigma=\Sigma_c=\Sigma(p_c+0)$, where  both $F(\Sigma_c,p_c)=0$ and  $\left.[dF(\Sigma,p)/d\Sigma]\right|_{\Sigma=\Sigma_c,p=p_c}=0$.
For every topology of the network of networks, we have square root singularities at the transition points: 
\bea
\Sigma-\Sigma_c\propto (p-p_c)^{1/2},
\eea
similarly to what happens in multiplex networks \cite{Son,Dorogovtsev}. 
Substituting the resulting $\Sigma$ back into Eq.~\ref{sigma} we find its dependence on $p$ for different $q$. This dependence is plotted in Fig.~\ref{f3} for the same networks of networks as in Fig.~\ref{f3}. Note that for networks of networks with Poissonian layers, $\sigma_q$ coincides with $S_q$, which is the fraction of nodes of a layer with superdegree $q$, belonging to the giant mutually connected component, see Eq.~(\ref{ps}). 
In the networks of networks in which the minimal superdegree $m$ of the supernetwork is greater or equal to 2, i.e. $m\geq2$,  the only viable solution for the order parameter is $\Sigma=0\ \forall p,c$ and any superdegree distribution $P(q)$. In the case 
of $m=0$ there are layers that are not interacting  ($q=0$) with other layers. These layers can be treated separately without losing the generality of the treatment. Therefore the only nontrivial case is the case of $m=1$.


\subsection{Partial interdependence}

In this section we consider an interesting variation of the configuration model.
In particular we assume that the superdegree adjacency matrix ${\bf a}$ has probability $P({\bf a})$ given by Eq.~(\ref{Pa}), while the probability of $P(\{s_{i,\alpha}\})$ is given by Eq.~(\ref{Ps}).
In addition, we assume that each node $(i,\alpha)$ is interdependent on the linked to it nodes in other layers only with probability $r$. 
In this situation 
the nodes in layer $\alpha$ with superdegree $q_{\alpha}=q$ can be interdependent with $n\in[0,q]$ other randomly chosen layers.
For this network of networks the message passing equations are given by  
\bea
&&\hspace*{-10pt}\sigma_{(i,\alpha)\to (j,\alpha)}\!=\!\!\!\!\!\!\prod_{(i^{\prime},\gamma)\in {\cal C}({i,\alpha})\setminus \alpha}\!\!\left\{\!s_{i^{\prime}\gamma}\!\!\left[\!{1}-\!\!\!\!\prod_{\ell\in N_{\gamma}(i)}\!\!(1{-}\sigma_{(\ell,\gamma)\to (i^{\prime},\gamma)})\!\right]\!\!\right\}
\nonumber 
\\[5pt]
&&~~~~~~~~~~~\times s_{i,\alpha}\left[1-\prod_{\ell\in N_{\alpha}(i)\setminus j}(1-\sigma_{(\ell,\alpha)\to (i,\alpha)})\right],\nonumber 
\eea
where ${\cal C}(i,\alpha)$ is the connected component of the nodes that can be reached by following only those of interlinks that imply 
interdependence between the nodes.
We 
indicate by $\sigma_q$ the average messages within a layer $\alpha$ of degree $q_{\alpha}=q$, obtaining the following equations: 
\bea
\sigma_{q}&=&p\sum_{s}\sum_{n=0}^q\binom{q}{n}r^n(1-r)^{q-n}P(s|n)B^{s-1}
\nonumber 
\\[5pt]
&&\times[1-G_1^k(1-\sigma_{q})].
\label{sqpr}
\eea
Here, $P(s|n)$ indicates the probability that a node $i$ in layer $\alpha$ with $q_{\alpha}=q$ and $n$ interdependent layers is in a connected component ${\cal C}(i,\alpha)$ 
of cardinality  $|{\cal C}(i,\alpha)|=s$. The quantity $B$ in Eq.~(\ref{sqpr}) is defined as
\bea
B=\left[\sum_{q'}\frac{q' P(q')}{\Avg{q}}p[1-G_0^k(1-\sigma_{q'})]\right].
\label{B2}
\eea
Similarly, the probability that a node $i$ in a layer $\alpha$ with superdegree $q_{\alpha}=q$ is in the mutually connected component $S_q=\Avg{S_{i,\alpha}}$ is given by  by the following expression: 
\bea
&&S_q=p\sum_{s}\sum_{n=0}^q\binom{q}{n}r^n(1-r)^{q-n})P(s|n)B^{s-1}
\nonumber 
\\[5pt]
&&~~~~\times[1-G_0^k(1-\sigma_{q})].
\label{Sqpr}
\eea
Proceeding in a similar way to 
the preceding paragraph we can 
derive 
the following equations determining the size of the mutually connected component when $M\gg 1$:  
\bea
&&\sigma_q=p(r\Sigma+1-r)^q[1-G_1^k(1-\sigma_q)] 
,
\nonumber
\\[5pt]
&&S_q=p(r\Sigma+1-r)^q[1-G_0^k(1-\sigma_q)] 
,
\nonumber
\\[5pt]
&&\Sigma =\left\{\sum_{q'}\frac{q' P(q')}{\Avg{q}}p[1-G_0^k(1-\sigma_{q'})]\right\}
\nonumber 
\\[5pt]
&&~~~\times \sum_{q'}\frac{q' P(q')}{\Avg{q}}(r\Sigma+1-r)^{q'-1}
.
\eea
Let us consider for simplicity the case in which each layer is formed by a Poisson network with average degree $\avg{k}=c$. Then the previous equations become
\bea
&&\sigma_q=S_q=p(r\Sigma+1-r)^q(1-e^{-c\sigma_q})
, 
\nonumber
\\[5pt]
&&\Sigma=\left\{\sum_{q'}\frac{q' P(q')}{\Avg{q}}p[1-e^{-c\sigma_{q'}}]\right\}
\nonumber
\\[5pt]
&&\times \sum_{q'}\frac{q' P(q')}{\Avg{q}}(r\Sigma+1-r)^{q'-1}.
\label{psp}
\eea
In particular, if 
the local supernetwork is regular, i.e. $P(q)=\delta(q,m)$, we have 
$\sigma_m=\sigma$  satisfying the following equation: 
\bea
\sigma=p\left[\frac{1}{2}\left(1-r+\sqrt{(1-r)^2+4r\sigma}\right)\right]^{q}(1-e^{-c\sigma})
.
\eea
In the more general case of an arbitrary distribution $P(q)$ the problem defined in Eqs.~(\ref{ps}) continues to have  a single order parameter given by $\Sigma$. 
The first equation in Eqs.~(\ref{psp}) has solution expressed in terms of the principal value of the Lambert function $W(x)$, 
namely 
\bea
&&\!\!\!\!\!\!\!\!\!\!\!\!\sigma_q=\frac{1}{c}\left[pc (r\Sigma+1-r)^q \phantom{e^{-pc (r\Sigma+1-r)^{q}}} \right.
\nonumber
\\[5pt]
&&\!\!\!\!\!\!\!\!\!\!\!\!~~~\left. +W\left(-pc(r\Sigma+1-r)^{q}e^{-pc (r\Sigma+1-r)^{q}}\right)\right]
.
\eea
Inserting this solution 
into the second equation 
in Eqs.~(\ref{psp}) 
we obtain the following equation for $\Sigma$: 
\bea
&&\!\!\!\!\!\!\Sigma=G_1^q(r\Sigma+1-r)\left[p +\sum_{q}\frac{q P(q)}{\Avg{q}}\frac{1}{c}(r\Sigma+1-r)^{-q}\right.
\nonumber 
\\[5pt]
&&\!\!\!\!\!\!~~\left.\times W\left(-pc(r\Sigma+1-r)^{q}e^{-pc (r\Sigma+1-r)^{q}}\right)\right]
.
\label{op2}
\eea
This equation can be written as $F_2(\Sigma,p,r)=0$. By imposing both $F_2(\Sigma,p,r)=0$ and $dF_2(\Sigma,p,r)/d\Sigma=0$ we can find the set of critical points $p=p_c$ where there 
are jumps 
of  $\Sigma(p)$. 

The  equation $dF(\Sigma,p,r)/d\Sigma=0$ reads as 
\bea
&&\!\!\!\!\!\!\!\!\!\!\!\!1/p=\left[\sum_{q}\frac{q P(q)}{\Avg{q}}(1-e^{-c\sigma_q})\right]
\nonumber 
\\[5pt]
&&\!\!\!\!\!\!\!\!\!\!\!\!~~~~~\times \left\{\sum_q \frac{q(q-1)P(q)}{\avg{q}}[r\Sigma+1-r]^{q-2}\right.
\nonumber 
\\[5pt]
&&\!\!\!\!\!\!\!\!\!\!\!\!~~~~~+\left. cp G_1^q(r\Sigma+1-r)\sum_{q<q_{max}} \frac{q^2P(q)}{\avg{q}}e^{-c\sigma_q}\right.
\nonumber 
\\[5pt]
&&\!\!\!\!\!\!\!\!\!\!\!\!~~~~~\left.\times \frac{[r\Sigma+1-r]^{q-1}(1-e^{-c\sigma_q})}{1-pc[r\Sigma+1-r]^qe^{-c\sigma_q}}\right\}
,
\eea
where $q_{max}=[-\log(pc)/\log(r\Sigma+1-r)]$.
 
\begin{figure}[t]
\begin{center}
\includegraphics[width=4.1in]{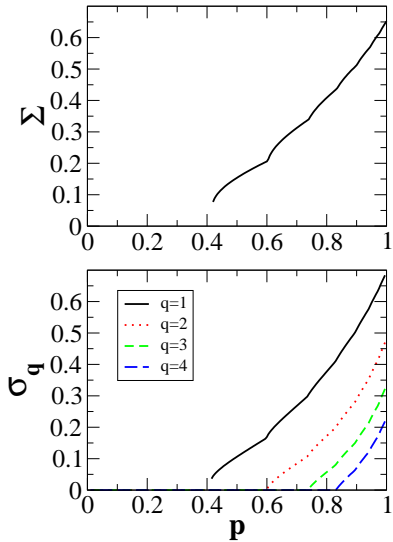}
\end{center}
\caption{(Color online) The order parameter $\Sigma$ and the  probabilities $\sigma_q$, that by following a link in a layer with superdegree $q$ we reach a node in the mutually connected giant component. 
$r=0.9$. The supernetwork is a scale-free network with the power-law degree distribution exponent $\gamma=2.9$, and each layer is a Poisson network with average degree $c=20$. The transition at with the layers with superdegree $q=1$ start to percolate is discontinuous, while all the other transitions are continuous.}
\label{fcd}
\end{figure}

Note that for $r\neq 1$, in contrast to the case of 
$r=1$, we can consider all values of the minimal superdegree $m>1$.
Interestingly, in addition to the abrupt phase transitions, this model displays also a set of continuous phase transitions, where the order parameter $\sigma_q$ acquires a nonzero value. These transitions occur at the points 
\bea
p_c=\frac{1}{c[1-r+r\Sigma(p_c,r)]^q}
.
\eea
These transitions are only stable for $r$ below some value $r_q$, namely  $r<r_q$,  and at $r=r_q$  they become discontinuous.
The value  $r_q$ can be obtained by solving 
the following system of equations:
\bea
&&\frac{1}{c(1-r+r\Sigma)^q}=p
,
\nonumber 
\\[5pt]
&&~~~~~~F_2(\Sigma,p,r)=0
,
\nonumber 
\\[5pt]
&&~~~~\frac{dF_2(\Sigma,p,r)}{d\Sigma}=0
.
\eea
In Fig.~\ref{fcd} for $r=0.9$, we show the order parameter $\Sigma$ of the network of networks and the probabilities $\sigma_q$ that by following a link in a layer with superdegree $q$ we reach a node in the mutually connected giant component. 
In this case, the phase transition at which the layers with $q=1$ start to percolate is discontinuous, while the other phase transitions are continuous. 

\subsection{Percolation properties of the network of networks with random supernetwork}

Thus the properties of the percolation transition in a random, locally tree-like supernetwork 
of layers with the same degree distribution $P(k)$ in the limit $M\gg1$ are the following:
\begin{itemize}

\item The size of the mutually connected giant component depends on the degree distribution $P(q)$ or the supernetwork and the degree distribution $P(k)$ of the individual layers.

\item
If the superdegree distribution $P(q)$ is heterogeneous, then there are multiple phase transitions at which a finite fraction of nodes in layers with different $q$ join the giant mutually connected component. For higher $q$, these transitions take place at higher values of the concentration $p$ of retained nodes. Therefore 
layers of high superdegree, corresponding to these transitions, 
are more fragile than layers with lower superdegree.

\item
The mutually connected components appears discontinuously if there is no partial interdependency.

\item
In the presence of a probability $r$ that an interlink indicates a real interdependency, there is a critical value $r=r_q$ such that as $p$ changes,  the percolation transition corresponding to the activation of layers with $q^{\prime}>q$ is continuous. 
\item 
The equations determining the size of the mutually connected component are different from the ones valid for a  network of networks in which the nodes can be linked only to their replica nodes, as, for example, in multiplex 
networks but 
coincide with equations for the configuration model of a network of networks.

\end{itemize}


\section{Conclusions}

In conclusion, here we provide a 
comprehensive treatment of the robustness of a network of networks in which each node of a layer can be interdependent on any node of the  connected layers in the supernetwork, see Fig.~\ref{f1}B. 
We show 
that the robustness of the network of networks is particularly dependent on the way the interlinks between the layer are connected.
We find that this model differs strongly from the network of networks in that each node has a replica node in each of the other layer and can be only interconnected to its replica nodes, see Fig.~\ref{f1}A.

The equations determining the size of the mutually connected component depend on the loop structure of the supernetwork. If the supernetwork is a tree, then the equations are the same as in the case of multiplex networks or in the case of a network of networks in the presence of replica nodes. If the supernetwork contains loops, then the equation determining the size of the mutually connected component changes significantly.
Here we provide as an example the treatment of a supernetwork formed by a single loop, and we show the 
strong effect of the loopy structure of the supernetwork on the robustness properties of the network.
For example, for a loop supernetwork formed by $M=3$ layers  
with replica nodes, we have the same transition as in a multiplex networks formed by $M=3$ layers. In contrast to that, 
we never observe a mutually connected component 
in the case of 
interdependencies between random nodes and of infinite number of nodes in each individual layer. 
Another major difference between the robustness of the networks of networks considered in this paper and the robustness of the network of networks with replica nodes, Fig.~\ref{f1}A, is that  in the latter case, the mutually connected component emerges simultaneously in all the layers, whereas if the interlinks can connect random nodes (not replica nodes) then the mutually connected component can have multiple phase transitions at which layer of higher superdegree begin percolate at different value of $p$, where $p$ is the probability that initially a random node is not damaged. 
Note that despite the fact that  the model presented in this work is very different from the configuration model (Fig.~\ref{f1}C) studied in \cite{BD2} the solution show significant similarities with it in the case of a random supernetwork formed by infinite number of layers. In the configuration model of network of networks \cite{BD2} each node of a given layer has a fixed number of superdegrees and each node can be interdependent on nodes in any random layer. Therefore the notion of the supernetwork is actually absent. In contrast to this, in the present work, there is a 
supernetwork,  
see Fig.~\ref{f1}B, and the nodes of each given layer are connected to random nodes of the interdependent layers. The similarity of the solutions of the two models exists when the supernetwork is random, and so 
the connected components ${\cal C}(i,\alpha)$ of the nodes $(i,\alpha)$ 
are random networks in both cases.
In particularly we show that 
the multiple phase transitions are observed when the supernetwork is an uncorrelated 
network with a given superdegree distribution. 
In this case the multiple phase transitions remain discontinuous and hybrid as long as there is no partial interdependence. In the presence of partial interdependence some of these transitions can become continuous, 
and 
we can observe coexistence of continuous and discontinuous phase transitions. 

Our work significantly expands the range of networks of networks with disordered interdependencies between nodes in different layers, in which the multiple phase transitions are observed. We suggest that this intriguing phenomenon should take place even in a wider class of networks of networks.

\acknowledgments

This work was partially supported by the FET proactive IP project MULTIPLEX 317532.


\end{document}